\documentstyle[epsfig,prl,twocolumn,aps]{revtex}
\begin{document}

\title{\bf Spin orbit induced torque in collinear spin valve structure and associated entropy}
\author{T.~P.~Pareek}
\address{Harish Chandra Research Institute, Chhatnag Road, Jhusi,
Allahabad - 211019, India}
\address{~
\parbox{14cm}{\rm
\medskip
We predict that due to spin-orbit(SO) interaction there is a torque 
even for the parallel configuration of spin valve structure (F1/2DEG/F2).
This torque arises due to spin orbit interaction. We develop a scattering theory for spin density matrix which allows us to study this effect quantitatively.
Further we show that the von-Neumann entropy associated with transport of polarization can show non-linear behavior as a function of absolute angle and 
oscillator behavior as a function of SO interaction strength.
}}
\maketitle
\narrowtext

Since the prediction of spin transfer torque phenomena ten year ago by 
Slonczewski \cite{slonz} and Berger \cite{berger}, considerable 
experimental evidence has been accumulated that a spin-polarized current can induce switching or precession of the films magnetization\cite{tsoi}\cite{wegrowe}\cite{sun}\cite{myers}
\cite{grollier}. In their pioneering work Slonczewski and Berger
predicted for spin valve geometry that for uniform and {\it non-collinear } magnetization in two ferromagnetic layers , magnetization of free layer will precess due to the spin transfer torque. This effect vanishes for collinear or parallel
magnetization since the spin transfer torque varies as sin($\theta$) where $\theta$ is the angle between the magnetization of two ferromagnetic layers
(see eq. 8 in Ref. \cite{berger}). 

On theoretical side a lot of work has been done finding new phenomena related to spin transfer torque for e.g. dynamic exchange coupling in magnetic bilayers\cite{bauer1}, spin echo \cite{bauer2} etc. On the other hand the phenomena
has been modeled by variety of ways including,combined ballistic and diffusive model\cite{guo}, first principal calculation\cite{bauer} 
microscopic spin transfer model
\cite{xiao}, and magnetization dynamics \cite{xi}.

All these studies were done in absence of spin-orbit (SO) interaction which is
rather important in low dimensional heterostructure.
In presence of SO interaction the situation differs dramatically. 
To elaborate this point we specifically consider a two dimensional 
heterostructure consisting of a two dimensional electron gas 
sandwiched between two ferromagnetic contacts (F1/2DEG/F2).
In such a system with a prominent source of SO interaction is asymmetric  confining potential which gives rise to Rashba spin-orbit (RSO) interaction and it magnitude can be varied by changing the external electric filed applied for 
confining the structure. We take the plane of 2DEG as {\it xy} plane and a normal to it as {\it z} axis which defines the coordinate system which will be used throughout this paper. In this frame the RSO interaction can be written as
\begin{equation}
H_{so}=\bf{B_{R}(k)\cdot\bbox{\sigma}},
\label{rso}
\end{equation} 
where  
${\bf B_{R}(k)}$=$ \lambda (k_{y}{\bf \hat{x}}-k_{x}{\bf \hat{y}})$ and $\bbox{\sigma}$ is a vector of Pauli matrices. Magnitude and direction of $\bf B_{R}(k)$ depends on instantaneous momentum vector, infact the direction is perpendicular to instantaneous wave vector.

When an electron is injected into 2DEG its spin precesses due to the RSO
given by eq.(1) as long as injected spin direction is not an eigen state of 
eq. (1). In hybrid structurer like F1/2DEG/F2 electrons are injected not only 
perpendicular to the interface rather over all possible angles. Hence the RSO will cause spin precession which implies that when electrons reaches detector
ferromagnet F2 it spin will develop a component perpendicular to the
magnetization of F2 which is parallel to the magnetization of F1. Hence this
perpendicular component will give rise to a torque on the magnetization F2 and vice versa. This torque arises due to the spin precession caused by RSO interaction and is none zero even for the parallel configuration of F1 and F2.

Further over the last decade , the physical characteristics of the purity and entanglement of quantum-mechanical states has been recognized as a central resources in various aspect of quantum computation \cite{nielsen} and spintronics. 
Since for spintronics operation spin-orbit interaction is
an integral part which entangles the orbital and spin degrees of freedom.
This entanglement affects the coherence of quantum state in a nontrivial way.
Therefore it is important to know how the
purity and entanglement are affected by the spin-orbit interaction. 
Motivated by this we
study in this paper, How does von-Neumann entropy (which is a
measure of entanglement and purity of state) depends on various parameters
of the problem?
As we will see that this is also related with the torque arising due to SO interaction. Therfore seemingly different
phenomena are related at fundamental level. Our theoretical formulation provides a unified way to addresses all these question in one setting.

In light of above discussion we develop a theory to study this
effect quantitatively and qualitatively. Since the state of ferromagnet is
not a coherent superposition of up and down electors rather it is a mixed state.
Hence a proper scattering theory should take this mixed character of ferromagnetic state into account\cite{nikolic}. This is done conveniently in density matrix formulation.
To this end let us assume that an electrons in F1 and F2 occupies states 
$|n,\alpha\rangle \equiv|n\rangle \otimes |\alpha\rangle$ where $n$ is channel index due to transverse confinement along y direction and $\alpha$ determines the spin state of the channels and can take tow values $\uparrow$ and $\downarrow$ respectively.
Let an electron be incident from F1 onto 2DEG from a state $\psi_{in}=|n,\alpha\rangle$ while the final sate in F2 is given as 
$\psi_{f}=T|n,\alpha\rangle=\sum_{m,\beta}|m,\beta\rangle \langle 
m,\beta|T|n,\alpha\rangle\equiv
\sum_{m,\beta} T_{m\,n}^{\beta\,\alpha}|m,\beta\rangle$. Here we have used the completeness relation $\sum_{m,\beta}|m,\beta\rangle\langle m,\beta|=1$ and $T_{m\,n}^{\beta\,\alpha}$ is transmission coefficient from state $|n,\alpha\rangle$ to state $|m,\beta\rangle$.
. The final state density matrix is given by
\begin{equation}
\rho_{f}^{n\,\alpha}=\frac{1}{N}|\psi_{f}\rangle\langle\psi_{f}|\equiv\sum_{{m\,\beta},{l\,\gamma}}T_{m\,n}^{\beta\,\alpha}T^{\dagger\,\alpha\,\gamma}_{n\,l}|m\,\beta\rangle\langle l\,\gamma|,
\label{rho_1}
\end{equation}
where N is normalization factor ensuring that $Tr(\rho)=1$.
Since we are interested in the spin degrees of freedom only hence we should take trace over 
orbital degree of freedom in eq. (\ref{rho_1}), which leads to the following 
equation for spin density matrix, {\it i.e.},
\begin{equation}
\rho_{f}^{n\,\alpha}=\frac{1}{N}\sum_{m}\sum_{\beta\,\gamma}
T_{m\,n}^{\beta\,\alpha}T^{\dagger\,\alpha\,\gamma}_{n\,m}
|\beta\rangle\langle\gamma|.
\end{equation}

Since the incident electrons are also originating in mixed state 
so the corresponding density matrix for the incident electron
is $\rho_{in}=n_{\alpha}|\alpha\rangle+n_{-\alpha}|-\alpha\rangle$, where $n_{\alpha}$ and $n_{-\alpha}$ is the number of up electrons and down electrons respectively.
The polarization P of ferromagnet is related to these as $P=(n_{\alpha}-n_{-\alpha})/(n_{\alpha}+n_{-\alpha})$. The fact that incident state is a mixed is incorporated into final state density matrix leading to the final state density matrix as,
\begin{equation}
\rho_{f}=n_{\alpha}\rho_{f}^{n\,\alpha}+n_{-\alpha}\rho_{f}^{n\,-\alpha}
\end{equation}
Once the final sate density matrix is known one can calculate the different 
component of transported polarization by taking respective trace of Pauli matrices {\it i.e.}, $P_{i}^{tr}=Tr(\rho_{f}\sigma_{i})$, where i={\it x, y or z}.

Von-Neumann entropy
for the final state density matrix is defined as,
\begin{equation}
S_{v}=-Tr(\rho_{f} \, ln(\rho_{f}))\equiv-\sum_{i}\lambda_{i}ln(\lambda_{i}),
\end{equation}
where $\lambda_{i}$ are the eigenvalues of the density matrix $\rho_{f}$.
From the definition it is clear that extremal values of $S_{v}$ are 0 and 1
for pure and completely mixed state respectively.

To obtain quantitative results we perform numerical simulation for a 
two dimensional lattice of 100$\times$100 sites.
We model the conductor on a square tight binding lattice
with lattice spacing {\it a} and we use the corresponding tight
binding model including spin orbit interaction given by
eq.(\ref{rso}) \cite{tribhu}. For the calculation of spin resolved
transmission coefficient , we use the recursive green function
method. Details of this can be found in
Ref. \cite{tribhu},\cite{tribhu1}. We fix the Fermi energy $E_{F}=1.0t$
above the bottom of band where $t$ is tight binding hopping parameter
we take $t$=1 as the unit of energy. The tight binding dimensionless 
SO parameter $\alpha=\lambda/t*a$ where $a$ is lattice spacing.
The numerical result presented takes the quantum effect
and multiple scattering into account.
For the model of disorder we take Anderson model,
where on-site energies are distributed randomly within [-U/2, U/2],
where U is the width of distribution. 

Fig. 1 show the component of the transported polarization {\it i.e.}, $P_{x}^{tr}$ , $P_{y}^{tr}$  and  $P_{z}^{tr}$ respectively for ballistic system
as a function of absolute angle $\theta$ of the magnetization direction of F1 and F2. We remind that since the system is two dimensional a natural frame
is defined by plane of 2DEG and a normal to the plane in right hand sens.
In this frame we rotate the magnetization of F1 and F2 along the direction
$(0, -\sin(\theta), \cos\theta)$ while keeping them always parallel {\it i.e.}
$\bf{P_{1}}=\bf{P_{2}}$=$P_{0}(0,-\sin(\theta),\cos(\theta))$. 
This implies that to begin with there is no component of polarization 
along {\it x} {\it i.e.} $P_{x}=0$ and it remains so in absence 
of spin-orbit interaction. This is confirmed by numerical calculation where
in Fig. 1 we have shown the results for zero SO interaction and as expected $P_{x}$ remains zero while $P_{y}$ and $P_{z}$ shows variation as $-\sin(\theta)$ and $\cos(\theta)$ as expected.
Further we see from Fig. 1 that $P_{x}^{tr} \neq 0$ for non zero strength of spin orbit interaction as well the other two components along {\it y} and {\it z} axis. This confirms our prediction made in introduction that
due to SO interaction, transported spin polarization will develop a component
perpendicular to the magnetization direction of F2. Hence the torque acting on 
F2 will be non zero even though F1 and F2 are parallel.
This is seen clearly in
Fig. 2 where torque is plotted as a function of absolute angle $\theta$ for ballistic($U=0.0$) and diffusive($U=0.5t$ and $U=1.0t$) case respectively. Dimensionless torque is defines as $Torque=\frac{1}{(\hbar\,I/\mid e \mid)}\mid{\bf P_{2}}\times {\bf P^{tr}}\mid\equiv P_{0}\sqrt(P^{tr\,2}_{x}+(P^{tr}_{y}\cos(\theta)+P^{tr}_{z}\sin(\theta)^2)$.
We notice that for ballistic case the torque is zero when $\bf{P_{1}}$ and $\bf{P_{2}}$ are parallel 
to {\it y} axis. This is so because injected electrons in this case are approximately an eigen state of RSO interaction given by eq.(\ref{rso}). While for the diffusive case torque is non zero for all angles $\theta$, since in diffusive case due to the scattering direction of $\bf{B_{R}}(k)$ is randomized hence probability of finding electron in eigenstate of SO interaction is reduced.

Left Panel of Fig. 3 shows magnitude of net transported polarization 
{\it i.e.},
$P^{tr}=\sqrt(P_{x}^{tr\,2}+P_{y}^{tr\,2}+P_{z}^{tr\,2})$ for ballistic system with varying SO interaction strength $\alpha$ . 
We notice that for $\theta=90$, $P^{tr}$ is almost equal to the value of 
polarization in the absence of spin-orbit interaction , shown as straight line.
This is consistent with the results of Fig. 2 where torque vanishes
for $\theta=90$. 
Fig. 3 (right panel) shows entropy
for ballistic system as a function of absolute angle $\theta$.
First we notice that $S_{v} < 1$ for $\alpha=0.0$ implying that
the states of Ferromagnetic leads F1 and F2 are a mixed state. Therefore
the use of density matrix scattering theory is essential and the only way to treat these mixed state situation.
Further as is seen entropy is minimum when $P^{tr}$ is
maximum reflecting the more pure character of state and is maximum when 
$P^{tr}$ is minimum reflecting the more mixed character of the state. Moreover
the entropy can be less compared to the vale for zero SO coupling for certain values of $\theta$ while corresponding $P^{tr}$ is large compared to the
value for zero spin orbit. {\it Implying that SO interaction 
has a polarizing effect and can in principal create a pure state from
mixed state \cite{tribhu_prl}. This is important since in general
any scattering process is reverse that is pure states evolve into mixed sates
and one looses the associated quantum character of states which is rather
important for spintronics and quantum computation}.

In Fig. 4 we study Torque and Entropy as a function of exchange splitting.
We see that torque remains zero when polarization of ferromagnet points
along {\it y} axis ($\theta=\pi/2$). This is in agreement with the results 
shown in Fig.2 and arises since the injected electrons are approximately in
the eigen state of SO interaction. While the torque along {\it z} axis
($\theta=0.0$) first increases with the increase of $\Delta$ and then drops
to zero at $\Delta/t=2.0$. At the same same point entropy also goes to zero
implying that the injected state is pure. Fig.5 confirms that at 
$\Delta/t=2.0$ state becomes pure since 
polarization goes to one. In other-words it is 
not entangled with
the orbital degrees of freedom. Since the state is decoupled from 
orbital degrees of freedom hence the torque arising due to spin-orbit
interaction which couples spin and orbital degrees of freedom goes to zero.
Therefore for pure state the torque arising due to SO interaction will be zero 
irrespective of along which direction it is pointing. Further this also
implies that torque is anisotropic and depends on absolute angle.

Fig. 6 and Fig. 7 shows Torque ,$P^{tr}$, and Entropy respectively along
{\it Z} axis as function of SO interaction $\alpha$. We see that all three
quantities shows oscillation as $\alpha$ is varied and the period of 
oscillation $\delta\alpha=0.03$,
which corresponds to precession angel of 
$\pi$ over length 100{\it a}. This length is the length of our 
systems. Hence these oscillation are due to spin precession caused by
SO interaction over the system length.

In conclusion we have shown that in collinear spin valve structure
a non zero torque arises due to the presence of SO interaction.
We have related this with the Von-Neumann entropy and studied their behaviors as
function of different parameters. We have formulated a spin density matrix
scattering theory which allows us to study all these seemingly different phenomena under one setting.

\vspace{0.5cm}
\begin{figure}
\begin{center}
\mbox{\epsfig{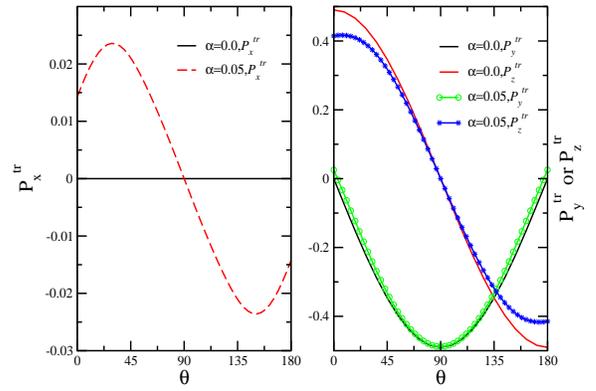}}
\end{center}
\caption{{Different comopnent of transported spin polarization as
a function of absolute angle $\theta$ for different values of dimensionless
SO interaction 
strength $\alpha$. Other parameters are $E_{F}$=1.0,and  
exchange splitting($\Delta$),  in Ferromagnet F1 and F2 is  given as
$2*\Delta/E_{F}=1.5$.}}
\label{Fig.1}
\end{figure}

\vspace{0.5cm}

\begin{figure}
\begin{center}
\mbox{\epsfig{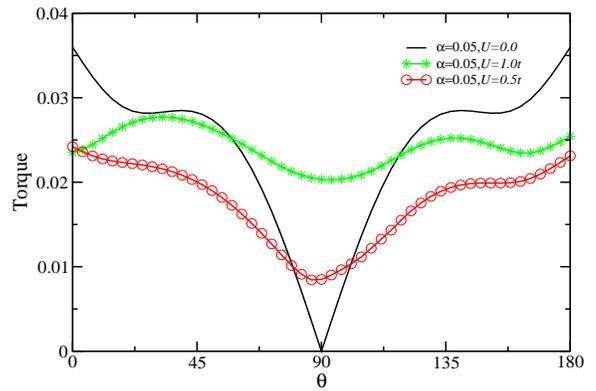}}
\end{center}
\caption{Dimensionless torque (defined in text) as a function of absolute angle $\theta$ for different values 
of dimensionless spin-orbit coupling $\alpha$.
Other parameters are same as for Fig. 2.}
\label{Fig.2}
\end{figure}

\begin{figure}
\begin{center}
\mbox{\epsfig{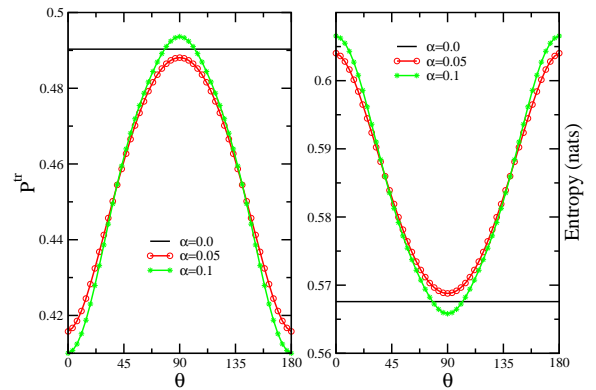}}
\end{center}
\caption{Transported polarization (left panel) and von-Neumann entropy as 
a function of angle $\theta$ for different values 
of dimensionless spin-orbit coupling $\alpha$.
Other parameters are same as for Fig. 2.}
\label{Fig.3}
\end{figure}

\begin{figure}
\begin{center}
\mbox{\epsfig{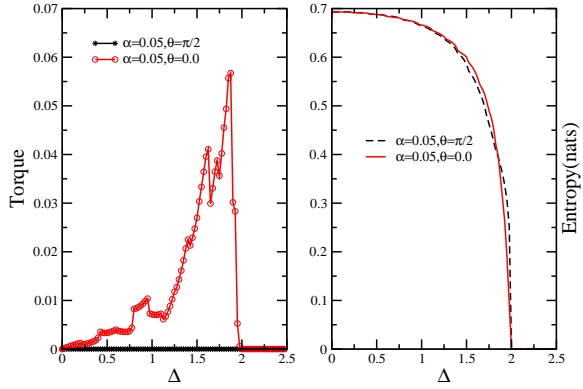}}
\end{center}
\caption{Dimensionless torque (defined in text) and Entropy along {\it y} and {\it z} axis
as function of exchange splitting $\Delta$.
Other parameters are same as for Fig. 1.}
\label{Fig.4}
\end{figure}

\begin{figure}
\begin{center}
\mbox{\epsfig{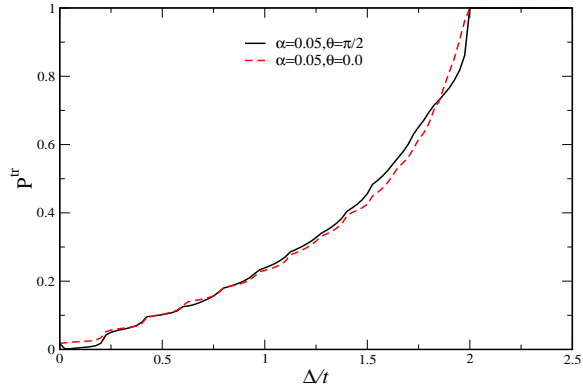}}
\end{center}
\caption{Polarization along {\it y} and {\it z} axis
as function of exchange splitting $\Delta$.
Other parameters are same as for Fig. 1.}
\label{Fig.5}
\end{figure}

\begin{figure}
\begin{center}
\mbox{\epsfig{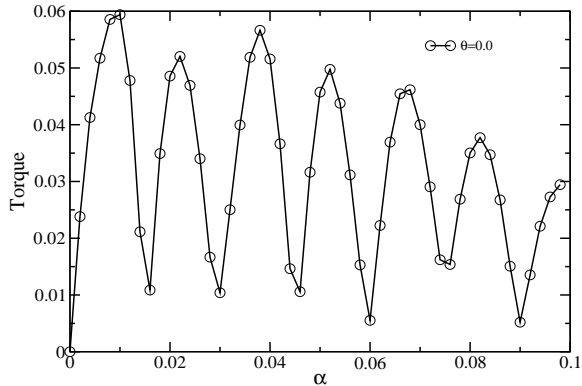}}
\end{center}
\caption{Dimensionless torque (defined in text) along {\it z} axis
as a function of dimensionless spin-orbit coupling $\alpha$.
Other parameters are same as for Fig. 1.}
\label{Fig.6}
\end{figure}

\begin{figure}
\begin{center}
\mbox{\epsfig{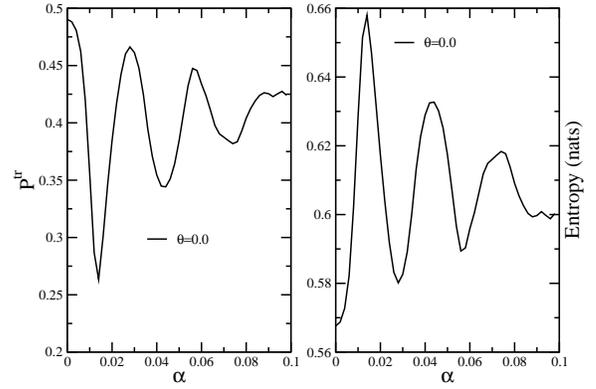}}
\end{center}
\caption{Transported polarization (left panel) and von-Neumann entropy as 
a function of  
of dimensionless spin-orbit coupling $\alpha$.
Other parameters are same as for Fig. 2.}
\label{Fig.7}
\end{figure}


\begin{thebibliography}{}
\bibitem{slonz} J.~C.~Slomczewski, J. Magn. Magn. Mater. {\bf 159} , L1 (1996).
\bibitem{berger} L. Berger, Phys. Rev. B. {\bf 54} , 9353 (1996).
\bibitem{tsoi} M. Tsoi, A.~G.~M. Jansen, J. ~Bass, W.~C.~Chiang, M.~Seck, V. ~Tsoi, and P. ~Wyder, Phys. Rev. Lett. {\bf 80}, 4281 (1998).
\bibitem{wegrowe}J. ~E.~Wegrowe, D.~Kelly, T.~Truong, Ph.~Guittienne, and J.~-Ph.~Ansemet, Europhys. Lett. {\bf 45}, 626 (1999).
\bibitem{sun} J.~Z.~Sun, J. Magn. Magn. Mater. {\bf 202}, 157 (1999).
\bibitem{myers} E.~B.~Myers, D.~C.~Ralph,J.~A.~Katine,R.~N.~Louie, and R.~A.~Buhrman, Science {\bf 285}, 867 (1999).
\bibitem{grollier} J.~Grollier, V.~Cross,A.~Hamzic,J.~M.~George, H.~Jaffres,A.~Fert, G.~Faini,J.~Ben~Youssef, and H.~Legall, Appl. Phys. Lett {\bf 78}, 3663(2001).
\bibitem{bauer1} B.~Heinrich,Y.~Tserkovnyak,G.~Woltersdorf,A.~Brataas, 
R.~Urban, and  G.~E.~W.~Bauer,Phys. Rev. Lett {\bf 90} , 187601-1(2003).
\bibitem{bauer2} A.~Brataas,G.~Zar\'and, Y.~Tserkovnyak,and  G.~E.~W.~Bauer
Phys. Rev. Lett {\bf 91} 166601-1 (2003).
\bibitem{guo} J. Guo and M.~B.~A.~Jalil, Phys. Rev. B. {\bf 71} 224408 (2005).
\bibitem{xi} H.~Xi,Y.~Shi,and K.~Z.~Gao, Phys. Rev. B. {\bf 71} 144418 (2005).
\bibitem{bauer} M.~Zwierzycki,Y.~Tserkovnyak, P.~J.~Kelly, A.~Brataas
and G.~E.~W.~Bauer, Phys. Rev. B {\bf 71} 0644420 (2005).
\bibitem{xiao} J.~xiao, A. Zangwillaw and M.~D.~Stiles, Phys. Rev. B. {\bf 72}
01446 (2005).
\bibitem{nielsen} M.~A.~Nielsen and I.~L.~Chuang, Quantum computation and Quantum information, Cambride Univ. Press , 2002.
\bibitem{nikolic} B.~K.~Nikolic and S.~Souma, Phys. Rev. B. {\bf 71}, 195328 (2005).
\bibitem{datta} S. Datta and B. Das, Appl. Phys. Lett.{\bf 56},
  665(1990).
\bibitem{tribhu}  T. P. Pareek and P. Bruno, Phys. Rev. B.{\bf 63}, 165424-1
(2001). P. Bruno Phys. Rev. Lett. {\bf 79}, 4593 (1997). 
\bibitem{tribhu1} T. P. Pareek, Phys. Rev. B. {\bf 66}, 193301 (2002).
T. P. Pareek and P. Bruno, Phys. Rev. B. {\bf 65}, 241305 (2002).

\bibitem{tribhu_prl} T. P. Pareek, Phys. Rev. Lett {\bf 92}, 076601 (2004).
\end{thebibliography}
\end{document}